\newcommand{\fix}[1]{{#1}} 
\documentclass[sensors,article,accept,moreauthors,pdftex]{Definitions/mdpinologo} 
\usepackage{graphicx}
\usepackage{lscape}
\usepackage{bbding}

\firstpage{1} 
\pubvolume{21}
\issuenum{18}
\articlenumber{6037}
\pubyear{2021}
\copyrightyear{2021}
\externaleditor{
} 
\datereceived{} 
\dateaccepted{} 
\datepublished{} 
\hreflink{\url{https://doi.org/10.3390/s21186037}} 

\Title{\fix{A Survey of Human Activity Recognition in Smart Homes Based on IoT Sensors Algorithms: Taxonomies, Challenges, and Opportunities with Deep Learning}}

\TitleCitation{A Survey of Human Activity Recognition in Smart Homes Based on IoT Sensors Algorithms: Taxonomies, Challenges, and Opportunities with Deep Learning}

\Author{{Damien Bouchabou} 
 $^{1,2,}$*\orcidA{}, Sao Mai Nguyen $^{1}$*\orcidB{}, Christophe Lohr $^{1}$\orcidC{}, Benoit LeDuc $^{2}$ and Ioannis Kanellos $^{1}$\orcidD{}}

\AuthorNames{Damien Bouchabou, Sao Mai Nguyen, Christophe Lohr, Benoit LeDuc, and Ioannis Kanellos}

\AuthorCitation{Bouchabou, D.; Nguyen, S.; Lohr, C.; LeDuc, B.; Kanellos, I.}

\address{%
$^{1}$ \quad IMT Atlantique, Brest, France; (damien.bouchabou, christophe.lohr, ioannis.kanellos)@imt-atlantique.fr\\
$^{2}$ \quad Delta Dore company, Bonnemain, France; (dbouchabou,bleduc)@deltadore.com\\
$^{3}$ \quad IMT Atlantique, Brest, France; nguyensmai@gmail.com}

\corres{Correspondence: damien.bouchabou@imt-atlantique.fr,nguyensmai@gmail.com)}

\abstract{Recent advances in Internet of Things (IoT) technologies and the reduction in the cost of sensors have encouraged the development of smart environments, such as smart homes. Smart homes can offer home assistance services to improve the quality of life, autonomy and health of their residents, especially for the elderly and dependent. To provide such services, a smart home must be able to understand the daily activities of its residents. Techniques for recognizing human activity in smart homes are advancing daily. But new challenges are emerging every day. In this paper, we present recent algorithms, works, challenges and taxonomy of the field of human activity recognition in a smart home through ambient sensors. Moreover, since activity recognition in smart homes is a young field, we raise specific problems, missing and needed contributions. But also propose directions, research opportunities and solutions to accelerate advances in this field.}


\begin{document}

\newcolumntype{Y}{>{\centering\arraybackslash}X}
\section{Introduction}

\begin{figure}[!b]
\centering
\includegraphics[width=\linewidth]{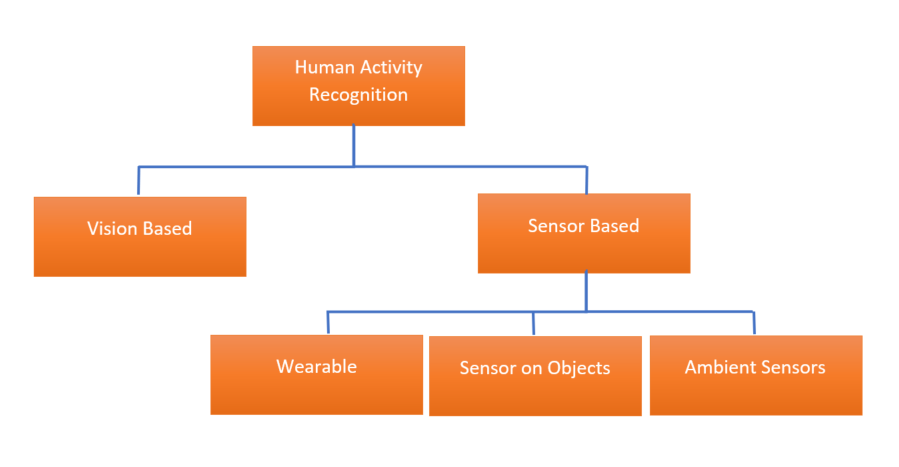}
\caption{Human Activity Recognition approaches}
\label{figure_1}
\end{figure}

With an ageing population, providing automated services to enable people to live as independently and healthily as possible in their own homes has opened up a new field of economics \cite{chan2008review}. Thanks to advances in the Internet of Things (IoT), the smart home is the solution being explored today to provide home services such as health care monitoring, assistance in daily tasks, energy management or security. A smart home is a house equipped with many sensors and actuators that can detect the opening of doors, the luminosity of the rooms, their temperature and humidity, \dots But also to control some equipment of our daily life as heating, shutters, lights or our household appliances. More and more of these devices are now connected and controllable at a distance. It is now possible to find in the houses, televisions, refrigerators, washing machines known as intelligent, which contain sensors and are controllable remotely. All these devices, sensors, actuators and objects can be interconnected through communication protocols.

In order to provide all of these services, a smart home must understand and recognise the activities of residents. To do so, the researchers are developing the techniques of Human Activity Recognition (HAR), which consists of monitoring and analysing the behaviour of one or more people to deduce the activity that is carried out. The various systems for HAR \cite{hussain2019different} can be divided into two categories \cite{dang2020sensor}: video-based systems and sensor-based systems (see Figure \ref{figure_1}). 

\subsection{Vision Based}
The vision based HAR uses cameras to track human behaviour and changes in the environment. This approach uses computer vision techniques, e.g. marker extraction, structure model, motion segmentation, action extraction, motion tracking. Researchers use a wide variety of cameras, from simple RGB cameras to more complex systems by fusion of several cameras for stereo vision or depth cameras able to detect the depth of a scene with infrared lights. 
Several survey papers about vision based activity recognition have been published \cite{beddiar2020vision, dang2020sensor}. Beddiar et al. \cite{beddiar2020vision} aims to provide an up-to-date analysis of vision based HAR-related literature and recent progress. 

However, these systems pose the question of acceptability. 
A recent study \cite{singh2018users} shows that the acceptability of these systems depends on users' perception of the benefits that such a smart home can provide. 
It also conditions their concerns about the monitoring and sharing the data collected. 
This study shows that older adults (ages 36 to 70) are more open to tracking and sharing data, especially if it is useful to their doctors and caregivers, while, younger adults (up to age 35) are rather reluctant to share information.
This observation argues for less intrusive systems, such as smart homes based on IoT sensors.

\subsection{Sensor Based}
HAR from sensors consists of using a network of sensors and connected devices, to track a person’s activity. They produce data in the form of a time series of state changes or parameter values. The wide range of sensors -- contact detectors, RFID, accelerometers, motion sensors, noise sensors, radar\dots -- can be placed directly on a person, on objects or in the environment. Thus the sensor-based solutions can be divided into three categories, respectively : Wearable \cite{ordonez2016deep}, Sensor on Objects \cite{li2016deep} and Ambient Sensor \cite{gomes2018intelligent}. 

Considering the privacy issues of installing cameras in our personal space, to be less intrusive and more accepted, sensor-based systems have dominated the applications of monitoring our daily activities \cite{chen2012sensor,hussain2019different}. 
Owing to the development of smart devices and Internet of Things, and the reduction of their prices, ambient sensor-based smart homes have become a viable technical solution which now needs to find human activity algorithms to uncover their potential. 

\subsection{Key Contributions}
While existing surveys \cite{aggarwal2014human,vrigkas2015review,hussain2019different,wang2019deep,chen2020deep} report past works in sensor-based HAR in general,  
we will focus in this survey on algorithms for human activity recognition in smart homes and its particular taxonomies and challenges for the ambient sensors, which we will develop in the next sections. Indeed, HAR in smart homes is a challenging problem because the human activity is something complex and variable from a resident to another. Every resident has different lifestyles, habits or abilities. The wide range of daily activities, the variability and the flexibility in how they can be performed require an approach that is scalable and must be adaptive.

\begin{figure}[!t]
\centering
\includegraphics[width=\linewidth]{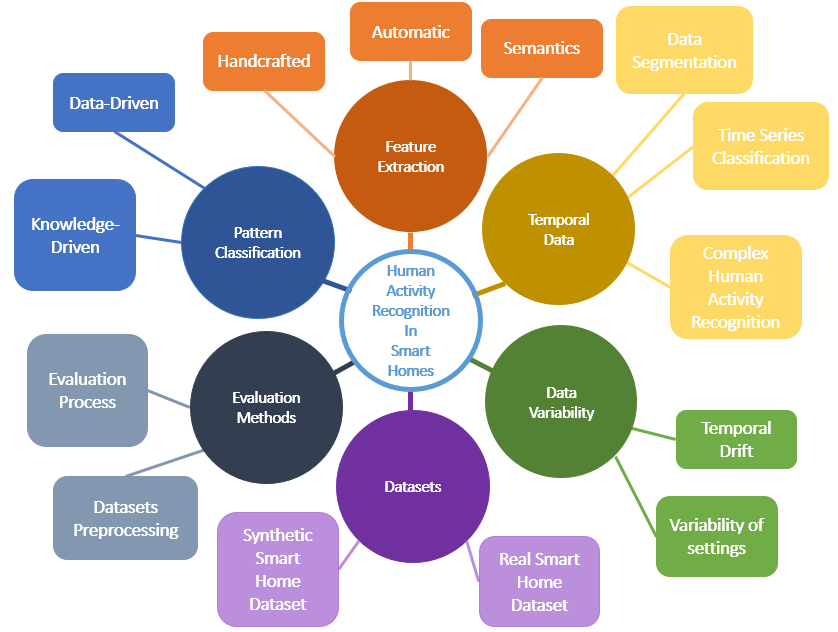}
\caption{Challenges for Human Activity Recognition in Smart Homes}
\label{fig:challenges}
\end{figure}

Many methods have been used for the recognition of human activity. However, this field still faces many technical challenges. Some of these challenges are common to other areas of pattern recognition  (sec. \ref{sec:classification}) and more recently on automatic features extraction algorithms (sec. \ref{sec:features}), such as computer vision and natural language processing, while some are specific to sensor-based activity recognition, and some are even more specific to the smart home domain. This field requires specific methods for real-life applications. The data have a specific temporal structure (sec. \ref{sec:temporal}) that needs to be tackled, and poses challenges in terms of data variability  (sec. \ref{sec:variability}) and availability of datasets (sec. \ref{sec:datasets}) but also specific evaluation methods (sec. \ref{sec:evaluation}). The challenges are summarised in fig. \ref{fig:challenges})

\fix{To carry out our review of the state of the art, we searched the literature for the latest advances in the field. We took the time to reproduce some works to confirm the results of works proposing high classification scores. In this study we were able to study and reproduce the work of \cite{liciotti_lstm,gochoo2018unobtrusive,yan2019using}, which allowed us to obtain a better understanding of the difficulties, challenges and opportunities in the field of HAR in smart homes.}

Compared with existing surveys, the key contributions of this work can be summarised as follows:

\begin{itemize}
\item We conduct a comprehensive survey of recent methods and approaches for human activity recognition in smart homes
\item We propose a new taxonomy of human activity recognition in smart homes in the view of challenges. 
\item We summarise recent works that apply deep learning techniques for human activity recognition in smart homes
\item We discuss some open issues in this field and point out potential future research directions.
\end{itemize}

\section{Pattern Classification}
\label{sec:classification}
Algorithms for Human Activity Recognition (HAR) in smart homes are first pattern recognition algorithms. The methods found in the literature can be divided into two broad categories: Data-Driven Approaches (DDA) and Knowledge-Driven Approaches (KDA). These two approaches are opposite. DDA uses user-generated data to model and recognize the activity. They are based on data mining and machine learning techniques. KDA uses expert knowledge and rule design. They use prior knowledge of the domain, its modeling and logical reasoning.

\subsection{Knowledge-Driven Approaches (KDA)}
In  KDA methods, an activity model is built through the incorporation of rich prior knowledge gleaned from the application domain, using knowledge engineering and knowledge management techniques.

KDA are motivated by real-world observations that involve activities of daily living and lists of objects required for performing such activities. In real life situations, even if the activity is performed in different ways, the number and objects type involved do not vary significantly. For example, the activity “brush teeth” contain actions involving a toothbrush, toothpaste, water tap, cup and towel. On the other hand, as humans have different lifestyles, habits, and abilities, they can perform various activities in different ways. For instance, the activity “make coffee” could be very different form one person to another.

KDA are founded upon the observations that most activities, specifically, routine activities of daily living and working, take place in a relatively circumstance of time, location and space. For example, brushing teeth are normally undertaken twice a day in a bathroom in the morning and before going to bed and involve the use of toothpaste and toothbrush. Thin implicit relationships between activities, related temporal and spatial context and the entities involved, provide a diversity of hints and heuristics for inferring activities.

The knowledge structure is modeled and represented through forms such as schemas, rules or networks. KDA modeling and recognition intends to make use of rich domain knowledge and heuristics for activity modeling and pattern recognition. Three sub approaches exist to use KDA, mining based approach \cite{perkowitz2004mining}, logic-based approach \cite{chen2008logical} and ontology based approach.

Ontology based approaches are the most commonly used, as ontological activity models do not depend on algorithmic choices. They have been utilized to construct reliable activity models. Chen et al. in \cite{chen2019human} have proposed an overview. 
Yamada et al. \cite{yamada2007applying} use ontologies to represent objects in an activity space. Their work exploits the semantic relationship between objects and activities. A teapot is used in an activity of tea preparation for example. This approach can automatically detect possible activities related to an object. It can also link an object to several representations or variability of an activity.

Chen et al. \cite{chen2009semantic,chen2009ontology,chen2011knowledge} constructed context and activity ontologies for explicit domain modeling.

KDA have the advantages to formalize activities and propose semantic and logical approaches. Moreover these representations try to be most complete as possible to overcome the activity diversity. However, the limitations of these approaches are the complete domain knowledge requirements to build activities models and the weakness in handling uncertainty and adaptability to changes and new settings. They need domain experts to design knowledge and rules.  New rules can break or bypass the previous rules. These limitations are partially solved in the DDA approaches.

\subsection{Data-Driven Approaches (DDA)}

The DDA for HAR include both supervised and unsupervised learning methods, which primarily use probabilistic and statistical reasoning. 
Supervised learning requires labelled data on which an algorithm is trained. After training, the algorithm is then able to classify the unknown data.

The DDA strength is the probabilistic modelling capacity. These models are capable of handling noisy, uncertain and incomplete sensor data. They can capture domain heuristics, e.g., some activities are more likely than others. They don't require a predefined domain knowledge. However DDA require much data and in the case of supervised learning, clean and correctly labelled data.

We observe that decision trees \cite{logan2007long}, conditional random fields \cite{vail2007conditional} or support vector machines \cite{fleury2009svm} have been used for HAR. Probabilistic classifiers such as the Naive Bayes classifier  \cite{brdiczka2008learning,van2007bayesian,cook2010learning} also showed good performance in learning and classifying offline activities when a large amount of training data is available. Sedkly et al. \cite{sedky2018evaluating} evaluated several classification algorithms such as AdaBoost, Cortical Learning Algorithm (CLA), Decision Trees, Hidden Markov Model (HMM), Multi-layer Perceptron (MLP), Structured Perceptron and Support Vector Machines (SVM). They reported superior performance of DT, LSTM, SVM and stochastic gradient descent of linear SVM. logistic regression or regression  functions. 

\subsection{Outlines}
To summarise, KDA propose to model activities following expert engineering knowledge, which is time consuming and difficult to maintain in case of evolution. DDA seems to yield good recognition levels and promises to be more adaptive to evolution and new situations. However, the DDA only yield good performance when given well-designed features as inputs. DDA needs more data and computation time than KDA, but the increasing number of datasets and the increasing computation power minimises these difficulties and allows today even more complex models to be trained, such as Deep Learning(DL) models which can overcome the dependency on input features. 

\section{Features Extraction} \label{har_feature}
\label{sec:features}

While the most promising algorithms for Human Activity Recognition in smart homes seem to be machine learning techniques, we describe how their performance depends on the features used as input. We describe how more recent machine learning has tackled this issue to generate automatically these features, and to propose end-to-end learning. We then highlight an opportunity to generate these features while taking advantage of the semantic of human activity.

\subsection{Handcrafted Features}
In order to recognize the activities of daily life in smart homes, researchers first used manual methods. These handcrafted features are made after segmentation of the dataset into explicit activity sequences or windows. In order to provide efficient activity recognition systems, researchers have studied different features \cite{chinellato2016feature}.

Initially Krishann et al. \cite{cook2013activity} and Yala et al. \cite{yala2015feature} proposed several feature vector extraction methods described below: baseline, time dependency, sensor dependency and sensor dependency extension. These features are then used by classification algorithms such as SVM or Random Forest to perform the final classification. 

\fix{Inspired by previous work, more recently Aminikhanghahi et al. \cite{aminikhanghahi2019enhancing} evaluate different types of sensor flow segmentations. But also listed different handmade features. Temporal features such as day of the week, time of day, number of seconds since midnight, or time between sensor transitions have been studied. Spatial features were also evaluated such as location. But also metrics such as, the number of events in the window or the identifier of the sensor appearing most frequently in the previous segments.}

\subsubsection{The Baseline Method} 
This consists of extracting a feature vector from each window. It contains the time of the first and last sensor events in the window, the duration of the window and a simple count of the different sensor events within the window. The size of the feature vector depends on the number of sensors in the datasets. For instance, if the dataset contains 34 sensors, the vector size will be 34 + 3. From this baseline reserchers upgrade the method to overcome different problems or challenges.

\subsubsection{The Time Dependence Method} 
This tries to overcome the problem of the sampling rate of sensor events. In most dataset, sensor events are not sampled regularly and the temporal distance of an event from the last event in the segment has to be taken into account. To do this, the sensors are weighted according to their temporal distance. The more distant the sensor is in time, the less important it is.

\subsubsection{The Sensor Dependency Method} 
This has been proposed to address the problem of the relationship between the events in the segment. The idea is to weight the sensor events in relation to the last sensor event in the segment. The weights are based on a matrix of mutual information between sensors, calculated offline. If the sensor appears in pair with the last sensor of the segment in other parts of the sensor flow, then the weight is high and respectably low otherwise.

\subsubsection{The Sensor Dependency Extension Method} This proposes to add the frequency of the sensor pair in the mutual information matrix. The more frequently a pair of sensors appears together in the dataset, the greater their weight.

\subsubsection{The Past Contextual Information Method} This is an extension of the previous approaches to take into account information from past sessions. The classifier does not know the activity of the previous segment. For example, the activity ``enter home'' can only appear after the activity ``leave home''. Naively the previous activity cannot be added to the feature vector. The algorithm might not be able to generalize enough. Therefore, Krishnan et al. \cite{cook2013activity} propose a two-part learning process. First the model is trained without knowing the previous activity. Then each prediction of activity in the previous segment is given to the classifier when classifying the current segment.

\subsubsection{The Latent Knowledge Method} This was recently proposed by Surong et al. \cite{yan2019using}. They improved these features by adding probability features. These additional features are learned from explicit activity sequences, in an unsupervised manner by a HMM and a Bayesian network. In their work, Surong et al. compared these new features with features extracted by deep learning algorithms such as LSTM and CNN. The results obtained with these unsupervised augmented features are comparable to deep learning algorithms. They conclude that unsupervised learning significantly improves the performance of handcrafted features.

\subsection{Automatic Features}

In the aforementioned works, machine learning methods for the recognition of human activity make use of handcrafted features. However, these extracted features are carefully designed and heuristic. There is no universal or systematic approach for feature extraction to effectively capture the distinctive features of human activities. 

Cook et al. \cite{cook2013activity} introduce few years ago an unsupervised method of discovering activities from sensor data based on a traditional machine learning algorithm. The algorithm searches for a sequence pattern that best compresses the input dataset. After many iterations the it reports the best patterns. These patterns are then clustered and given to a classifier to perform the final classification.

In recent years, deep learning has flourished remarkably by modelling high-level abstractions from complex data \cite{pouyanfar2018survey} in many fields such as computer vision, natural language processing, and speech processing \fix{\cite{ordonez2016deep}. Deep learning models have the end-to-end learning capability to automatically learn high-level features from raw signals without the guidance of human experts, which facilitates their wide applications.} Thus, researchers used Multi Layer Perceptron (MLP) in order to carry out the classification of the activities \cite{fang2014human,irvine2020neural}. However, the key point of deep learning algorithms is their ability to learn features directly from the raw data in a hierarchical manner, eliminating the problem of crafty approximations of features. They can also perform the classification task directly from their own features. Wang et al. \cite{wang2019deep} presented a large study on deep learning techniques applied to HAR with the sensor-based approach. Here only the methods applied to smart homes are discussed.

\subsubsection{Convolutional Neural Networks (CNN)}
Works using Convolutional Neural Networks (CNN) has been carried out by the researchers. The CNN have demonstrated their strong capacity to extract characteristics in the field of image processing and time series. 
The CNN have two advantages for the HAR. First, they can capture local dependency, i.e. the importance of nearby observations correlated with the current event. And, they are scale invariant in terms of step difference or event frequency. In addition, they are able to learn a hierarchical representation of the data. There are two types of CNN: 2D CNN for image processing and 1D CNN for sequence processing.

Gochoo et al. \cite{gochoo2018unobtrusive} have transformed activity sequences into binary images in order to use 2D CNN-based structures. Their work showed that this type of structure could be applied to the HAR. In an extension, Gochoo et al. \cite{tan2018multi} propose to use coloured pixels in the image to encode new sensor information about the activity in the image. Their extension proposes a method to encode sensors such as temperature sensors, which are not binary, as well as the link between the different segments. Mohmed et al. \cite{mohmed2020employing} adopt the same strategy but convert activities into greyscale images. The grey value is correlated to the duration of sensor activation. The AlexNet structure \cite{krizhevsky2012imagenet} is then used for the feature extraction part of the images. Then, these features are used with classifiers to recognize the final activity. 

Singh et al. \cite{singh2017convolutional} used a CNN 1D-based structure on raw data sequences for their high feature extraction capability. Their experiments show that the CNN 1D architecture achieves similar hight results.

\subsubsection{Autoencoder Method}
Autoencoder is an unsupervised artificial neural network that learns how to efficiently compress and encode data then learns how to reconstruct the data back from the reduced encoded representation to a representation that is as close to the original input as possible. Autoencoder, by design, reduces data dimensions by learning how to ignore the noise in the data. Researchers have explored this possibility because of the strong capacity of Autoencoders to generate the most discriminating features. The reduced encoded representation created by the Autoencoder contains the features that allow to dicriminate the activities.

Wang et al. in \cite{wang2016human} apply a two-layer Stacked Denoising Autoencoder (SDAE) to automatically extract unsupervised meaningfully features. The input of the SDAE are feature vectors extracted from 6 second time windows without overlap. The feature vector size is the number of sensors in the dataset.
They compared two features forms: binary representation and numerical representation. The numerical representation method, records the number of firing of a sensor during the time window, while the binary representation method sets to one the sensor value if this one fired in the time window.
Wang et al. then use a dense layer on top of the SDAE to fine-tune this layer with the labeled data to perform the classification. Their method outperforms machine learning algorithms on the Van Kasteren Dataset \cite{van2011human} with the two features representations.

Ghods et al. \cite{ghods2019activity2vec} proposed a method, Activity2Vec to learn an activity Embedding from sensor data.  They used a Sequence-to-Sequence model (Seq2Seq) \cite{sutskever2014sequence} to encode and extract automatic features from sensors. The model trained as an Autoencoder, to reconstruct the initial input sequence in output. Ghods et al. validate the method with two datasets from HAR domain, one was composed of accelerometer and gyroscope signals from a smartphone and another one that contained smart sensor events.
Their experiment shows that the Activity2Vec method generates good automatic features. They measured the intra-class similarities with handcrafted and Activity2Vec features. It appears that for the first dataset (smartphone HAR) intra-class similarities are smallest with the Activity2Vec encoding. Conversely, for the second dataset (smart sensors events), the intra-class similarities are smallest with handcrafted features.

\subsection{Semantics}
Previous work has shown that deep learning algorithms such as Autoencoder or CNN are capable of extracting features but also of performing classification. They thus allow the creation of so-called end-to-end models. But these models  do not translate semantics representing the relationship between activities, as ontologies could represent these relationships. But in recent years, researchers in the field of Natural Language Processing (NLP) have developed techniques of word embedding and the language model for deep learning algorithms to understand not only the meaning of words but also the structure of phases and texts. A first attempt to add NLP word embedding to deep learning has shown a better performance in daily activity recognition in smart homes \cite{Bouchabou2021I2WDLHAR}. Moreover, the use of the semantics of the HAR domain may allow the development of new learning techniques for quick adaptation such as zero-shot learning, which is developed in sec. \ref{sec:variability}. 

\subsection{Outlines} 
All handcrafted methods for extracting features have produced remarkable results in many HAR applications. These approaches assume that each dataset has a set of features that are representative, allowing a learning model to achieve the best performance. However, handcrafted features require extensive pre-processing. This is time consuming and inefficient because the dataset is manually selected and validated by experts. This reduces adaptability to various environments. This is why HAR algorithms must automatically extract the relevant representations.

Methods based on deep learning allow better and higher quality features to be obtained from raw data. Moreover, these features can be learned for any dataset. They can be processed in a supervised or unsupervised manner, for example windows labelled or not with the name of the activity. In addition, deep learning methods can be end-to-end, i.e. they extract features and perform classification. \fix{Thanks to deep learning, great advances have been made in the field of NLP. It allows to represent words, sentences or texts thanks to models, structures and learning methods. These models are able to interpret the semantics of words, to contextualize them, to make prior or posterior correlations between words and thus to increase their performance in terms of sentence or text classification. Moreover, these models are able to automatically extract the right features to accomplish their task. The NLP and HAR domains in smart homes both process data in the form of sequences. In smart homes, sensors generate a stream of events. This stream of events is sequential and ordered like words in a text. Some events are correlated to earlier or later events in the stream. This stream of events can be segmented into sequences of activities. These sequences can be similar to sequences of words or sentences. Moreover, semantic links between sensors or types of sensors or activities may exist \cite{yamada2007applying}. We suggest that some of these learning methods or models can be transposed to deal with sequences of sensor events. We think in particular of methods using attention or embedding models.} 

However, these methods developed for pattern recognition might not be sufficient to analyse these data which are in fact temporal series.

\section{Temporal Data}
\label{sec:temporal}
In a smart home, sensors record the actions and interactions with the residents' environment over time. These recordings are the logs of events that capture the actions and activities of daily life. Most sensors only send their status when there is a change in status, to save battery power and also to not overload wireless communications. In addition, sensors may have different triggering times. This results in scattered sampling of the time series and irregular sampling. Therefore, recognizing human activity in a smart home is a pattern recognition problem in time series with irregular sampling, unlike recognizing human activity in videos or wearables.

In this section, we describe literature methods for segmentation of the sensor data stream in a smart home. These segmentation methods provide a representation of sensor data for human activity recognition algorithms. We highlight the challenges of dealing with the temporal complexity of human activity data in real use cases.

\subsection{Data Segmentation}

As in many fields of activity recognition, a common approach consists in segmenting the data flow. Then, using algorithms to identify the activity in each of these segments. Some methods are more suitable for real-time activity recognition than others. Real time is a necessity to propose reactive systems. In some situations, it is not suitable to recognise activities several minutes or hours after they occur, for example in case of emergencies such as fall detection. \fix{Quigley et al. \cite{quigley2018comparative} have studied and compared different windowing approaches.}

\subsubsection{Explicit Windowing (EW)}
This consists of parsing the data flow per activity \fix{\cite{cook2013activity,yala2015feature}}. Each of these segments corresponds to one window that contain a succession of sensor events belonging to the same activity. This window segmentation depends on the labelling of the data. In the case of absence of labels it is necessary to find the points of change of activities. The algorithms will then classify these windows by assigning the right activity label. This approach has some drawbacks. First of all, it is necessary to find the segments corresponding to each activity in case of unlabelled data. In addition, the algorithm must use the whole segment to predict the activity. It is therefore not possible to use this method in real time.

\subsubsection{Time Windows (TW)} The use of TW consists in dividing the data stream into time segments with a regular time interval. This approach is intuitive, but rather favorable to the time series of sensors with regular or continuous sampling over time. This is a common technique with wearable sensors such as accelerometers and gyroscopes. One of the problems is the selection of the optimal duration of the time interval. If the window is too small, it may not contain any relevant information. If it is too large, then the information may be related to several activities, and the dominant activity in the window will have a greater influence on the choice of the label. Van Kasteren et al. \cite{van2011activity} determined that a window of 60s is a time step that allows a good classification rate. This value is used as a reference in many recent works \cite{medina2018ensemble,hamad2019efficient,hamad2020joint,hamad2021dilated}. \fix{Quigley et al. \cite{quigley2018comparative} show that TW achieves a high accuracy but does not allow to find all classes.}

\subsubsection{Sensor Event Windows (SEW)} A SEW divides the stream via a sliding window into segments containing an equal number of sensor events. Each window is labeled with the label of the last event in the window. The sensor events that precede the last event in the window define the context of the last event. This method is simple but has some drawbacks. This type of window varies in terms of duration. It is therefore impossible to interpret the time between events. However, the relevance of the sensor events in the window can be different depending on the time interval between the events \cite{krishnan2014activity}. Furthermore, because it is a sliding window, it is possible to find events that belong to the current and previous activity at the same time. In addition, The size of the window in number of events, as for any type of window, is also a difficult parameter to determine. This parameter defines the size of the context of the last event. If the context is too small, there will be a lack of information to characterize the last event. However, if it is too large, it will be difficult to interpret. A window of 20 - 30 events is usually selected in the literature \cite{aminikhanghahi2019enhancing}.

\subsubsection{Dynamic Windows (DW)} DW uses a non-fixed window size unlike the previous methods. It is a two-stage approach that uses an offline phase and an online phase \cite{al2016windowing}. In the offline phase, the data stream is split into EW. From the EW, the "best-fit sensor group" is extracted based on rules and thresholds. Then, for the online phase, the dataset is streamed to the classification algorithm. When it identifies the "best-fit sensor group" in the stream, the classifier associates the corresponding label with the given input segment. Problems can arise if the source dataset is not properly annotated. Quigley et al. \cite{quigley2018comparative} have shown that this approach is inefficient for modeling complex activities. Furthermore, rules and thresholds are designed by experts, manually, which is time consuming.

\subsubsection{Fuzzy Time Windows (FTW)} \fix{FTW were introduced in the work of Medina et al.\cite{medina2018ensemble}. This type of window was created to encode multi-varied binary sensor sequences i.e. one series per sensor. The objective is to generate features for each sensor series according to its short, medium and long term evolution for a given time interval. As for the TW, the FTW segments the signal temporarily. However, unlike other types of window segmentation, FTW use a trapezoidal shape to segment the signal of each sensor. The values defining the trapezoidal shape follow the Fibonacci sequence, which resulted in good performance during classification. The construction of a FTW is done in two steps. First, the sensor stream is resampled by the minute, forming a binary matrix. Each column of this matrix represents a sensor and each row contains the activation value of the sensor during the minute i.e. 1 if the sensor is activated in the minute or 0 otherwise. For each sensor and each minute a number of FTW is defined and calculated. Thus each sensor for each minute is represented by a vector translating its activation in the current minute but also its past evolution. The size of this vector is related to the number of FTW.  This approach allowed to obtain excellent results for binary sensors. 
Hamand et al. \cite{hamad2019efficient} have proposed an extension of FTW by adding FTW using the future data of the sensor in addition to the past information. The purpose of this complement is to introduce a delay in the decision making of the classifier.} The intuition is that relying only on the past is not enough to predict the right label of activity and that in some cases delaying the recognition time allows to make a better decision. To illustrate with an example, if a binary sensor deployed on the front door generates an opening activation, the chosen activity could be ``the inhabitant has left the house''. However, it may happen that the inhabitant opens the front door only to talk to another person at the entrance of the house and comes back home without leaving. Therefore, the accuracy could be improved by using the activation of the following sensors. It is therefore useful to introduce a time delay in decision making. The longer the delay, the greater the accuracy. \fix{But a problem can appear if this delay is too long, indeed the delay prevents real time. While a long delay may be acceptable for some types of activity, others require a really short decision time in case of an emergency, i.g the fall of a resident. Furthermore, FTW are only applicable to binary sensors data and do not allow the use of non-binary sensors. However, in a smart home the sensors are not necessarily binary e.g. humidity sensors.}

\subsubsection{Outlines}
The table summarizes and categorizes the different segmentation techniques detailed above.
\begin{table}[h!]
\centering
\resizebox{0.6\textwidth}{!}{%
\begin{tabularx}{\textwidth}{|Y|Y|Y|Y|Y|Y|Y|Y|}
\hline
\textbf{Segmentation type} & \textbf{Usable for Real Time} & \textbf{Require resamplig} & \textbf{Time representation} & \textbf{Usable on raw data} & \textbf{Capture long term dependencies}         & \textbf{Capture dependence between sensors} & \textbf{\# steps} \\
\hline
EW                & No                   & No                & No                  & Yes                & only inside the sequence               & Yes                                & 1            \\
\hline
SEW               & Yes                  & No                & No                  & Yes                & depends of the size                    & Yes                                & 1            \\
\hline
TW                & Yes                  & Yes               & Yes                 & Yes                & depends of the size                    & No                                 & 1            \\
\hline
DW                & Yes                  & No                & No                  & Yes                & only inside the pre segmented sequence & Yes                                & 2            \\
\hline
FTW               & Yes                  & Yes               & Yes                 & Yes                & Yes                                      & No                                 & 2     \\
\hline   
\end{tabularx}%
}
\caption{Summary of segmentation methods}
\label{tab:segmentation_resume}
\end{table}

\subsection{Time Series Classification}

The recognition of human activity in a smart home is a problem of pattern recognition in time series with irregular sampling. Therefore more specific machine learning for sequential data analysis have also proven efficient for HAR in smart homes. 

Indeed, statistical Markov models such as Hidden Markov Models \cite{cook2009assessing,cook2010learning} and their generalisation,  Probabilistic graphical models as Dynamic Bayesian Networks \cite{philipose2004inferring} can model spatiotemporal information. In the deep learning framework, they have been implemented as Recurrent Neural Networks (RNN). RNN show today a stronger capacity to learn features and represent time series or sequential multi-dimensional data.

RNN are designed to take a series of inputs with no predetermined limit on size. RNN remembers the past and its decisions are influenced by what it has learnt from the past. RNN can take one or more input vectors and produce one or more output vectors and the output(s) are influenced not just by weights applied on inputs like a regular neural network, but also by a hidden state vector representing the context based on prior input(s)/output(s). So, the same input could produce a different output depending on previous inputs in the series. But RNN suffers from the long-term dependency problem \cite{bengio1994learning}. To avoid this problem two RNN variations have been proposed, the Long Short Term Memory (LSTM) \cite{hochreiter1997long} and Gated Recurrent Unit (GRU) \cite{cho2014learning}, which is a simplification of the LSTM. 

Liciotti et al. in \cite{liciotti_lstm} studied different LSTM structures on activity recognition. They showed that the LSTM approach outperforms traditional HAR approaches in terms of classification score without using handcrafted features, as LSTM can generate features that encode the temporal pattern. The higher performance of LSTM was also reported in \cite{singh2017recurent} in comparison traditional machine learning techniques (Naive Bayes, HMM, HSMM and Conditional Random Fields).  Likewise, Sedkly et al \cite{sedky2018evaluating} reported that LSTM perform better than  AdaBoost, Cortical Learning Algorithm (CLA), Hidden Markov Model or Multi-layer Perceptron or Structured Perceptron. Nevertheless, the LSTM still have limitations, and their performance is not significantly higher than decision Trees, SVM and stochastic gradient descent of linear SVM. logistic regression or regression  functions. Indeed LSTM still have difficulties to find the suitable time scale to balance between long-term temporal dependencies and short term temporal dependencies. A few works have attempted to tackle this issue.  Park et al. \cite{park2018deep} used a structure using multiple LSTM layers with residual connections and an attention module. Residual connections reduce the gradient vanishing problem, while the attention module marks important events in the time series. To deal with variable time scales, Medina-Quero et al. \cite{medina2018ensemble} have combined the LSTM with a fuzzy window to process the HAR in real time, as fuzzy windows can automatically adapt the length of its time scale. 
With accuracies lower than 96\%, these refinements still need to be consolidated and improved. 
 
\subsection{Complex Human Activity Recognition}
Besides, these sequential data analysis algorithms can only process simple, primitive activities, and can not yet deal with complex activities.\fix{A simple activity is an activity that consists of a single action or movement such as walking, running, turning on the light, opening a drawer. A complex activity is an activity that involves a sequence of actions potentially involving different interactions with objects, equipment or other people. For example, cooking.}  

\subsubsection{Sequences of sub-activities}
Indeed, activities of daily living  are not \emph{micro actions} such as gestures that are carried out the same way by all individuals. 
Activities of daily living that our smart homes want to recognise can be on the contrary seen as sequences of micro actions, which we can call \emph{compound actions}.  These sequences of micro actions generally follow a certain pattern, but there are no strict constraints on their compositions or the order of micro actions. This idea of compositionality was implemented by an ontology hierarchy of context-aware activities: a tree hierarchy of activities link each activity to its sub-activities \cite{Hong2009PMC}. Another work proposed a method to learn this hierarchy: as the  Hidden  Markov  Model  approach  is  not  well  suited  to process long sequences, an extension of  HMM  called  Hierarchical  Hidden  Markov  Model was proposed in \cite{Asghari2019} to encode multilevel dependencies in terms of time and follow a hierarchical structure in their context. To our knowledge, there have not been extensions of such hierarchical systems using deep learning, but hierarchical LTSM using two-layers of LSTM to tackle the varying composition of actions for  HAR based on videos proposing  \cite{Devanne2019ICSC} or using tow hidden layers in the LSTM for HAR using wearables \cite{Wang2020CSSP} can constitute inspirations for HAR in smart home applications. Other works in video-based HAR  proposed to automatically learn a stochastic grammar describing the hierarchical structure of complex activities from annotations acquired from multiple annotators \cite{Tayyub2018IWCACV}. 

The idea of these HAR algorithms is to use the context of a sensor activation, either by introducing multi-timescale representation to take into account longer term dependencies or by introducing context-sensitive information to channel the attention in the stream of sensor activations. 

The latter idea can be developed much further by taking advantage of the methods developed by the field of natural language processing, where texts also have a multi-level hierarchical structure, where the order of words can vary and where the context of a word is very important. Embedding techniques such as ELMo \cite{peters2018deep} based on LSTM or more recently BERT \cite{devlin2018bert} based on Transfomers \cite{vaswani2017attention} have been developed to handle sequential data while handling long-range dependencies through context-sensitive embeddings. 
These methods model the context of words to help the processing of long sequences. Applied to HAR, they could model the context of the sensors and their order of appearance. Taking inspiration from \cite{Tayyub2018IWCACV,Bouchabou2021I2WDLHAR}, we can draw a parallel between NLP and HAR: a word is apparent to a sensor event, a micro activity composed of sensor events  is apparent to a sentence, a compound activity composed of sub-activities is a paragraph.  The parallel between word and sensor events has led to the combination of word encodings with deep learning to improve the performance of HAR in smart homes in \cite{Bouchabou2021I2WDLHAR}. 

\subsubsection{Interleave and Concurrent Activities}
Human activities are often carried out in a complex manner. Activities can be carried out in an interleave or concurrent manner. An individual may alternately cook and wash dishes, or cook and listen to music simultaneously, but could just as easily cook and wash dishes alternately while listening to music. The possibilities are infinite in terms of activity scheduling. However, some activities seem impossible to see appearing in the dataset and could be anomalous, such as cooking while the individual sleeps in his room.

Researchers are working on this issue. Modeling this type of activity is becoming complex. But it could be modeled as a multi label classification problem. Safyan and .al \cite{safyan2019ontology} have explored this problem using ontology. Their approach uses a semantic segmentation of sensors and activities. This allows the model to relate the possibility that certain activities may or may not occur at the same time for the same resident. Li et al. \cite{li2017concurrent} exploit a CNN-LSTM structure to recognise concurrent activity with multimodal sensors.

\subsubsection{Multi-user Activities}

Moreover, monitoring the activities of daily living performed by a single resident is already a complex task. The complexity increases with several residents. The same activities become more difficult to recognise. On the one hand, a group, a resident may interact to perform common activities. In this case, the activation of the sensors reflects the same activity for each resident in the group. On the other hand, everyone can perform different activities simultaneously. This produces a simultaneous activation of the sensors for different activities. These activations are then merged and mixed in the activity sequences. An activity performed by one resident is a noise for the activities of another resident.

Some researchers are interested in this problem. As with the problem of recognising competing activities. The multi-resident activity recognition problem is a multi-label classification problem \cite{alhamoud2016activity}. Tran et al. \cite{tran2018multi} tackled the problem using a multi-label RNN. Natani et al. \cite{natani2021sequential} studied different neural network architectures such as MLP, CNN, LSTM, GRU, or hybrid structures to evaluate which structure is the most efficient. The hybrid structure that combines a CNN 1D and a LSTM is the best performing one.

\subsection{Outlines} 

\fix{A number of algorithms have been studied for HAR in smart homes. The Table \ref{tab:algorithms} show a summary and comparison of recent HAR methods in smart homes.}

\begin{table}[ht!]
\centering
\resizebox{0.6\textwidth}{!}{%
\begin{tabularx}{\textwidth}{|Y|Y|Y|Y|Y|Y|Y|Y|}
\hline
\textbf{ref}& \textbf{Segmentation} & \textbf{Data representaion} & \textbf{Encoding}& \textbf{Feature type} & \textbf{Classifier}& \textbf{Dataset}& \textbf{Real-time} \\ 
\hline
\cite{liciotti_lstm} & EW & Sequence & Integer sequence (one integer for each possible sensors activations) & Automatic & Uni LSTM, Bi LSTM, Cascade LSTM, Ensemble LSTM, Cascade Ensemble LSTM & CASAS \cite{cook2012casas}: Milan, Cairo, Kyoto2, Kyoto3, Kyoto4 & No \\ 
\hline
\cite{singh2017recurent}           & TW                    & Multi-channel               & Binary matrix                                                        & Automatic             & Uni LSTM                                                              & Kasteren \cite{van2011human}                                               & Yes                \\ \hline
\cite{park2018deep}                & EW                    & Sequence                    & Integer sequence (one integer for each sensor Id)                    & Automatic             & Residual LSTM, Residual GRU                                            & MIT \cite{tapia2004activity}                                               & No                 \\ \hline
\cite{medina2018ensemble}          & FTW                   & Multi-channel               & Real values matrix(computed values inside each FTW)                  & Manual                & LSTM                                                                  & Ordonez \cite{ordonez2013activity}, CASAS A \& CASAS B \cite{cook2012casas} & Yes                \\ \hline
\cite{gochoo2018unobtrusive}       & EW + SEW              & Multi-channel               & Binary picture                                                       & Automatic             & 2D CNN                                                                & CASAS \cite{cook2012casas}: Aruba                                          & No                 \\ \hline
\cite{hamad2020joint}              & FTW                   & Multi-channel               & Real values matrix(computed values inside each FTW)                  & Manual                & Joint LSTM + 1D CNN                                                   & Ordonez \cite{ordonez2013activity}, Kasteren \cite{van2011human}           & Yes                \\ \hline
\cite{singh2017convolutional}      & TW                    & Multi-channel               & Binary matrix                                                        & Automatic             & 1D CNN                                                                & Kasteren \cite{van2011human}                                               & Yes                \\ \hline
\cite{wang2020activities}          & TW                    & Multi-channel / Sequence    & Binary matrix, Binary vector, Numerical vector, Probability vector   & Automatic / Manual    & Autoencoder, 1D CNN, 2D CNN, LSTM, DBN                                & Ordonez \cite{ordonez2013activity}                                         & Yes                \\ \hline
\cite{aminikhanghahi2019enhancing} & SEW                   & Sequence                    & Categorical values                                                   & Manual                & Random Forest                                                         & CASAS \cite{cook2012casas}: HH101-HH125                                    & Yes                \\ \hline
\end{tabularx}%
}
\caption{Summary and comparison of activity recognition methods in smart homes}
\label{tab:algorithms}
\end{table}

LSTM shows excellent performance on the classification of irregular time series in the context of a single resident and simple activities. However, human activity is more complex than this. And challenges related to the recognition of concurrent, interleaved or idle activities offer more difficulties. Previous cited works did not take into account these type of activities. Moreover, people rarely live alone in a house. This is why even more complex challenges are introduced, including the recognition of activity in homes with multiple residents. These challenges are multi-class classification problems and still unsolved.

In order to address these challenges, activity recognition algorithms should be able to segment the stream for each resident. Techniques in the field of image processing based on Fully Convolutional Networks \cite{long2015fully} as U-Net \cite{ronneberger2015u} allow to segment the images. These same approaches can be adapted to time series \cite{perslev2019u} and can constitute inspirations for HAR in smart home applications.

\section{Data Variability}
\label{sec:variability}
Not only are real human activities complex, the application of human activity recognition in smart homes for real-use cases also faces issues causing a discrepancy between training and test data. The next subsections detail the issues inherent to smart homes : the temporal drift of the data and the variability of settings.

\subsection{Temporal drift}

Smart homes through their sensors and interactions with residents collect data on the behaviour of residents. Inital training data is the portrait of the activities performed at the time of registration. A model is generated and trained using this data. Over time, the behaviour and habits of the residents may change. The data that is now captured is no longer the same as the training data. It corresponds to a  time drift as introduced in \cite{schlimmer1986incremental}. This concept means that the statistical properties of the target variable, which the model is trying to predict, evolve over time in an unexpected way. A shift in the distribution between the training data and the test data. 

To accommodate this drift, algorithms for HAR in smart homes should incorporate \emph{life-long learning} to continuously learn and adapt to changes in human activities from new data as proposed in \cite{Thrun1998}. Recent works in life-long learning incorporating deep learning as reviewed in \cite{Parisi2019} could help tackle this issue of temporal drift. In particular, one can imagine that an interactive system can from time to time request labelled data to users to continue to learn and adapt. Such algorithms have been developed under the names of interactive reinforcement learning or active imitation learning in robotics. 
In \cite{Duminy2021AS} they allowed the system to learn micro and compound actions while minimising the number of requests for labelled data by choosing when, what information to ask, and to whom to ask for help. Such principles could inspire a smart home system to continue to adapt its model while minimising user intervention and optimising his intervention by pointing out the missing key information.

\subsection{Variability of Settings}

Beside these long-term evolutions, the data from one house to another are also very different, and the model learned in one house is hardly applicable in another because of the change in house configuration; sensors equipment; and families' compositions and habits. 
Indeed, the location, the number and the sensors type of smart homes can influence activity recognition systems performances.
Each smart homes can be equipped in different ways and have different architecture in terms of sensors, room configuration, appliance\dots Some can have a lot of sensors, multiple bathrooms, or bedrooms and contain multiple appliances. When others can be smaller as a single apartment, where sensors can be fewer and have more overlaps and noisy sequences. Due to this difference in house configurations, a model that optimised in the first smart homes could perform poorly in another. This issue could be solved by collecting a new dataset for each new household to train the models anew, however this is  costly as explained in sec. \ref{sec:datasets}. 

Another solution is to adapt the models learned in a household to another. Transfer learning methods have recently been developed to allow pre-trained deep learning models to be used with different data distributions, as reviewed in \cite{Weiss2016JD}. Transfer learning using deep learning  has been successfully applied to time series classification as reviewed in \cite{Ismail-Fawaz2018}. For activity recognition, Cook et al. \cite{Cook2013KIS} reviewed the different types of knowledge that could be transferred  in  traditional machine learning. These methods can be updated with deep learning algorithms and by benefiting from recent advances in transfer learning for deep learning. 
Furthermore, adaptation to new settings have recently been improved by the development of meta-learning algorithms. Their goal  is to train a model on a variety of learning tasks, so it can solve new learning tasks using only a small number of training samples. This field has seen recent breakthroughs as reviewed in \cite{Hospedales2020}, which has never been applied yet to HAR. Yet, the peculiar variability of data of HAR in smart homes can only benefit from such algorithms.  


\section{Datasets} \label{har_datasets}
\label{sec:datasets}
Datasets are key to train, test and validate activity recognition systems. Datasets were first generated in laboratories. But these records don't allow enough variety and complexity of activities and were not real enough. To overcome these issues public datasets were created from recordings in real homes with volunteer residents. In parallel to being able to compare in the same condition and on the same data, some competitions were created such as Evaluating AAL Systems Through Competitive Benchmarking - AR (EvAAL-AR) \cite{gjoreski2015competitive} or UCAmI Cup \cite{espinilla2018ucami}.

However, the production of datasets is a tedious task and recording campaigns are difficult to manage. They require volunteer actors and apartments or houses equipped with sensors. In addition, data annotation and post-processing take a lot of time. Intelligent home simulators have been developed as a solution to generate datasets.

This section presents and analyzes some real and synthetic data sets in order to understand the advantages and disadvantages of these two approaches.

\subsection{Real Smart Home Dataset}
A variety of public real homes datasets exist \cite{tapia2004activity,van2011human,cook2012casas,alemdar2013aras,cumin2017dataset}. De-la-Hoz et al. \cite{de2018sensor} provides an overview of sensor-based datasets used in HAR for smart homes. They compiled documentation and analysis of a wide range of datasets with a list of results and applied algorithms. But such dataset production implies some problems as: sensors type and placement, variability in term of user profile or typology of dwelling and the annotation strategy.

\subsubsection{Sensor Type and Positioning Problem}

When acquiring data in a house, it is difficult to choose the sensors and their numbers and locations. It is important to select sensors that are as minimally invasive as possible in order to respect the privacy of the volunteers \cite{alemdar2013aras}. No cameras or video recordings were used. The majority of sensor-oriented smart home datasets use so-called low-level sensors. These include infrared motion sensors (PIR), magnetic sensors for openings and closures, pressure sensors placed in sofas or beds, sensors for temperature, brightness, monitoring of electricity or water consumption \dots

The location of these sensors is critical to properly capture activity. Strategic positioning allows to accurately capture certain activities. e.g. a water level sensor in the toilet to capture toilet usage or a pressure sensor under a mattress to know if a person is in bed. There is no precise method or strategy for positioning and installing sensors in homes. CASAS \cite{cook2012casas}  researchers have proposed and recommended a number of strategic positions. However, some of these strategic placements can be problematic in terms of evolution. It is possible to imagine that during the life of a house the organization or use of its rooms changes e.g. if a motion sensor is placed above the bed to capture its use. But if the bed is moved to a different place in the room, then the sensor will no longer be able to capture this information. In the context of a dataset and the use of the dataset to validate the algorithms, this constraint is not important. But it becomes important in the context of real applications to evaluate the resilience of algorithms, which must continue to function in case of loss of information.

In addition to positioning, it is important to choose enough sensors to cover a maximum of possible activities. The number of sensors can be very different from one dataset to another. For example, the MIT dataset \cite{tapia2004activity} uses 77 and 84 sensors for each of these apartments. The Kasteren dataset \cite{van2011human} uses between 14 and 21 sensors. ARAS \cite{alemdar2013aras} has apartments with 20 sensors. Orange4Home \cite{cumin2017dataset} is based on an apartment equipped with 236 sensors. This difference can be explained by the different types of dwellings but also by the number and granularity of the activities that we want to recognize. Moreover some dataset are voluntarily over-equipped. There is still no method or strategy to define the number of sensors installed according to an activity list.

\subsubsection{Profile and Typology Problem}
\fix{It is important to take into account that there are different typologies of houses: apartment, house, with garden, with floors, without floor, one or more bathrooms, one or more bedrooms... These different types and variabilities of houses lead to difficulties such as: the possibility that the same activity takes place in different rooms. That the investment in terms of number of sensors can be more or less important. Or that the network coverage of the sensors can be problematic. 
For example, Alerndar et al. \cite{alemdar2013aras} faced a problem of data synchronization.} One of their houses required two sensor networks to cover the whole house. They must synchronized the data for dataset needs.
It is therefore necessary that the datasets can propose different house configurations in order to evaluate the algorithms in multiple configurations. Several dataset with several houses exist \cite{tapia2004activity,van2011human,van2011human,alemdar2013aras}. CASAS \cite{cook2012casas} is one of them, with about 30 several houses configurations. These datasets are very often used in the literature \cite{de2018sensor}. However the volunteers are mainly elderly people, and cover several age groups is important. A young resident does not have the same behavior as an older one. The Orange4Home dataset \cite{cumin2017dataset} cover the activity of a young resident. The number of residents is also important. The activity recognition is more complex in the case of multiple residents. This is why several datasets cover this field of research also \cite{cook2012casas,alemdar2013aras,van2011human}.

\subsubsection{Annotation Problem}

Dataset annotation is something essential for supervised algorithm training. When creating these datasets it is necessary to deploy strategies to enable this annotation. Such as journal \cite{van2011human}, smartphone applications \cite{cumin2017dataset}, personal digital assistant (PDA) \cite{tapia2004activity}, Graphical User Interface (GUI) \cite{alemdar2013aras} or voice records to annotate the dataset \cite{van2011human}. 

As these recordings are made directly by volunteers, they are asked to annotate their own activities. For the MIT dataset \cite{tapia2004activity}, residents used a PDA to annotate their activities. Every 15 minutes, the PDA beeped to prompt residents to answer a series of questions to annotate their activities, however, several problems were encountered with this method of user self-annotation. However, several problems were encountered with this method of self-annotation by the user, such as some short activities not being entered, errors in label selection, or omissions. A post-annotation based on the study of a posteriori activations was necessary to overcome these problems, thus potentially introducing new errors. In addition, this annotation strategy is cumbersome and stressful because of the frequency of inquiries. It requires great rigor from the volunteer and at the same time interrupts activity execution by pausing it when the information is given. These interruptions reduce the fluidity and natural flow of activities.

Van Kasteren et al. \cite{van2011human} proposed another way of annotating their data. The annotation was also done by the volunteers themselves, but using voice through a Bluetooth headset and a journal. This strategy allowed the volunteers to be free to move around and not need to create breaks in the activities. This allowed for more fluid and natural sequences of activities. The Diary allowed the volunteers to complete some additional information when wearing a helmet was not possible. However, wearing a helmet all day long remains a constraint.

The volunteers of the ARAS dataset \cite{alemdar2013aras} used a simple Graphical User Interface (GUI) to annotate their activities. Several instances were placed in homes to minimize interruptions in activities and avoid wearing an object such as a helmet all day long. Volunteers were asked to indicate only the beginning of each activity. It is assumed that residents will perform the same activity until the next start of the activity. This assumption reflects a bias that sees human activity as a continuous stream of known activity.

\subsection{Synthetic Smart Home Dataset}
The cost to build real smart homes and the collection of datasets for such scenarios is expensive and sometimes infeasible for many projects. Measurements campaigns should include a wide variety of activities and actors. It should be done with sufficient rigor to obtain qualitative data. Moreover, finding the optimal placement of the sensors \cite{helal2010scalable}, finding appropriate participants \cite{helal2011persim,mendez2009simulating} and the lack of flexibility \cite{armac2007simulation,fu2011configurable} makes the dataset collection difficult. For these reasons researchers imagined smart homes simulation tools \cite{alshammari2017openshs}.

These simulation tools can be categorized into two main approaches, model-based \cite{lee2015persim} and interactive \cite{synnott2014creation}, according to Synnott et. al. \cite{synnott2015simulation}. The model-based approach uses predefined models of activities to generate synthetic data. In contrast, the interactive approach relies on having an avatar that can be controlled by a researcher, human participant or simulated participant. Some hybrid simulator as OpenSH \cite{alshammari2017openshs} can combines advantages from both interactive and model-based approaches. In addition, smart homes simulation tool can be focusing on the dataset generation or data visualization. Some simulation tools provide multi-resident or fast forwarding to accelerate the time during execution.

These tools allow you to quickly generate data and visualize it. But the capture of activities can be unnatural and not noisy. Some uncertainty may be missing

\subsection{Outlines}
All these public datasets, synthetic or real, are useful and allow evaluating processes. Both, show advantages and drawbacks. \fix{The Table \ref{tab:datasets} details some datasets from the literature, resulting from the hard work of the community.}

\begin{table}[ht!]
\centering
\resizebox{0.6\textwidth}{!}{%
\begin{tabularx}{\textwidth}{|Y|Y|Y|Y|Y|Y|Y|Y|Y|}
\hline
\textbf{Ref}& \textbf{Multi resident} & \textbf{Resident type} & \textbf{Duration}& \textbf{Sensor type}& \textbf{\# of Sensors} & \textbf{\# of Activity} & \textbf{\# of Houses} & \textbf{Year} \\
\hline
\cite{van2011human}        & No             & Eldery        & 12-22 days & Binary         & 14-21        & 8             & 3           & 2011\\
\hline
\cite{alemdar2013aras}     & Yes            & Young         & 2 months   & Binary         & 20           & 27            & 3           & 2013\\
\hline
\cite{cumin2017dataset}    & No             & Young         & 2 weeks    & Binary, Scalar & 236          & 20            & 1           & 2017\\
\hline
\cite{cook2012casas}       & Yes            & Eldery        & 2-8 months & Binary, Scalar & 14-30        & 10-15         & >30         & 2012\\
\hline
\cite{ordonez2013activity} & No             & Eldery        & 14-21 days      & Binary         & 12           & 11            & 2           & 2013\\
\hline
\cite{tapia2004activity}   & No             & Eldery        & 2 weeks    & Binary, Scalar & 77-84        & 9-13          & 2           & 2004\\
\hline
\end{tabularx}%
}
\caption{Example of real datasets of the literature}
\label{tab:datasets}
\end{table}

Real datasets such as Orange4Home \cite{cumin2017dataset} provide a large sensor set. That can help to determine which sensors can be useful for which activity. CASAS \cite{cook2012casas} propose many houses or apartment configurations and topologies with elderly people. Which allows evaluating the adaptability to house topologies. ARAS \cite{alemdar2013aras} propose younger people and multi-residents' livings. Useful to validate the noisy resilience and segmentation ability of the activity recognition system. The strength of real datasets is their variability, and their representativeness in number and execution of activities. 
But sensors can be placed too strategically and wisely choose to cover some specific kinds of activities. In some datasets PIR sensors are used as a grid or installed as a checkpoint to track residents trajectory. Strategic placement, a large number of sensors or the choice of a particular sensor is great to help algorithms to infer knowledge but are not the real ground truth.

Synthetic datasets allow to quickly evaluate different configuration sensors and topologies. In addition they can produce large amounts of data without real setup or volunteer subjects. The annotation is more precise compared to real dataset methods (diary, smartphone apps, voice records).

But activities provided by synthetic datasets are less realistic in terms of execution rhythm and variability. Every individual has its own rhythm in terms of action duration, interval or order. The design of the virtual smart homes can be a tedious task for a non-expert designer. \fix{Moreover, no synthetic datasets are publicly available. Only some dataset generation tools as OpenSH \cite{alshammari2017openshs} are available.}

Today, even if smart sensors become cheaper and cheaper, real houses are not equipped with a wide range of sensors as it can be found in datasets. It is not realistic to find an opening sensor on a kitchen cabinet. Real homes contains PIR to monitor wide areas with the security system. Temperature sensors to control the heat. More and more air qualitative or luminosity sensors can be found. Some houses are now equipped with smart lights or smart plugs. Magnetic sensors can be found on external openings. And now, some houses provide general electrical and water consumption. These datasets are not representative of the actual home sensor equipment.

Another issue as shown above is the annotation. Supervised algorithms needs qualitative labels to learn correct features and classify activities. Residents' self-annotation can produce errors and lack of precision. Post processing to add annotations, adds uncertainty as they are always based on hypothesis, such as every activity is performed sequentially. But the human activity flow is not always sequential. Very few datasets provide concurrent or interleaved activities. Moreover every dataset proposes its own taxonomy for annotations. Even if synthetic datasets try to overcome annotation issues, 

This section demonstrates the difficulty of providing a correct evaluation system or dataset. And the work already provided by all the scientific community is excellent. Thanks to this amount of work, it is possible to, in certain conditions, evaluate activity recognition systems. 

\fix{However, there are several areas of research that can be explored to help the field progress more quickly. A first possible research axis for data generation is, the generation of data from video games. Video games constitute a multi-billion dollar industry, where developers put great effort into build highly realistic worlds. Recent works in the field of semantic video segmentation consider and use video games to generate datasets in order to train algorithms \cite{richter2016playing,richter2017playing}. Recently Roitberg et al. \cite{roitberg2021let} studied a first possibility using a commercial game by Electronic Arts (EA) " The Sims 4", a daily life simulator game, to reproduce the video Toyota Smarthome dataset \cite{das2019toyota}. The objective was to evaluate and train HAR algorithms from video produced by a video game and compare them to the original dataset. This work showed promising results. An extension of this work could be envisaged in order to generate datasets of sensor activity traces. Moreover, every dataset proposes its own taxonomy. Some are inspired by medical works such as, Katz et al. work \cite{katz1983assessing}, to define a list of basic and necessary activities. However there is no proposal for a hierarchical taxonomy e.g. cook lunch and cook dinner are children activities of cook. Or taxonomy taking into account concurrent or parallel activities. The suggestion of a common taxonomy for datasets is a research axis to be studied in order to homogenize and compare algorithms more efficiently.}

\section{Evaluation Methods}
\label{sec:evaluation}
In order to validate the performance of the algorithms, the researchers use datasets. But learning the parameters of a prediction function and testing it on the same data is a methodological error: a model that simply repeats the labels of the samples it has just seen would have a perfect score but could not predict anything useful on data that is still invisible. This situation is called overfitting. To avoid it, it is common practice in a supervised machine learning experiment to retain some of the available data as a dataset for testing. Several methods exist in the field of machine learning and deep learning. For the problem of HAR in smart houses, some of them have been used by researchers. 

The evaluation of these algorithms is not only related to the use of these methods. It depends on the methodology but also on the datasets on which the evaluation is based. It is not uncommon that preprocessing is necessary. However this preprocessing can influence the final results.  This section highlights some of the biases that can be induced by preprocessing the datasets as well as the application and choice of certain evaluation methods.

\subsection{Datasets Preprocessing}

\subsubsection{Unbalenced Datasets Problem}
Unbalanced datasets pose a challenge because most of the machine learning algorithms used for classification have been designed assuming an equal number of examples for each class. This results in models with poor predictive performance, especially for the minority class. This is a problem because, in general, the minority class is larger and the problem is therefore more sensitive to classification errors for the minority class than for the majority class. To get around this problem some researchers will rebalance the dataset. By removing classes that are too little represented. By randomly removing examples for the most represented classes \cite{gochoo2018unobtrusive} These approaches allow to increase the performance of the algorithms but do not allow to represent the reality. 

Within the context of the activities of daily life, certain activities are performed more or less often during the course of the days. A more realistic approach is to group activities under a new, more general label. Ex: ``preparing breakfast'', ``preparing lunch'', ``preparing dinner'', ``preparing a snack'', can be grouped under the label ``preparing a meal''. Therefore, activities that are less represented but semantically close can be used as parts of example. This can allow fairer comparisons between datasets if the label names are shared. Liciotti et al. \cite{liciotti_lstm} have adopted this approach to compare several datasets between them. One of the drawbacks is the loss of granularity of activities.

\subsubsection{The Other Class Issue}

In the field of HAR in smart houses, it is very frequent that a part of the dataset is not labeled. Usually the label ``Other'' is assigned to these unlabeled events. The class ``Other'' generally represents 50\% of the dataset \cite{liciotti_lstm, yan2019using}. This makes it the most represented class in the dataset and unbalances the dataset. Furthermore, the ``Other'' class may represent several different activity classes or simply something meaningless. Some researchers choose to suppress this class, judged to be over-represented and containing too many random sequences. Others prefer to remove it from the training phase and therefore from the training set. However, they keep it in the test set in order to evaluate the system in a more real-life environment \cite{yala2015feature}.  Yala et al. \cite{yala2015feature} evaluated performance with and without the ``Other'' class and showed that this choice has a strong impact on the final results.

However, being able to dissociate this class opens perspectives. Algorithms able to isolate these sequences could propose to the user to annotate them in the future in order to discover new activities.

\subsubsection{Labelling Issue}
As noted above, the datasets for the actual houses are labeled by the residents themselves, via a logbook or graphical user interface. They are then post-processed by the responsible researchers. However, it is not impossible that some labels may be missing as in the CASAS Milan dataset \cite{cook2012casas}. Table \ref{table:6} presents an extract from the Milan dataset where labels are missing. However, events or days are duplicated, i.e. same timestamp, same sensor, same value, same activity label. A cleaning of the dataset must be considered before the algorithms are formed.  Obviously, depending on the quality of the labels and data, the results will be different. Indeed some occurrence of classes could be artificially increased or decreased. Some events could be labeled ``Other'' even though they actually belong to a defined activity. In this case the recognition algorithm could label this event correctly but it would appear to be confused with another class in the confusion matrix.

\begin{table}[ht!]
\centering
{\scriptsize
\begin{tabular}{llllp{2cm}}  
\hline
Date & Time & Sensor ID & Value & Label \\
\hline
2010-01-05	&   08:25:37.000026	&   M003	&   OFF &   \\
2010-01-05	&   08:25:45.000001	&   M004	&   ON	&   Read begin\\
\dots & \dots & \dots & \dots & \dots \\
2010-01-05	&   08:35:09.000069	&   M004	&   ON &   \\	
2010-01-05	&   08:35:12.000054	&   M027	&   ON &   \\	
2010-01-05	&   08:35:13.000032	&   M004	&   OFF &   (Read should end) \\	
2010-01-05	&   08:35:18.000020	&   M027	&   OFF &   \\	
2010-01-05	&   08:35:18.000064	&   M027	&   ON &   \\	
2010-01-05	&   08:35:24.000088	&   M003	&   ON &   \\	
2010-01-05	&   08:35:26.000002	&   M012	&   ON &   (Kitchen Activity should begin)\\	
2010-01-05	&   08:35:27.000020	&   M023	&   ON &   \\	
\dots & \dots & \dots & \dots & \dots \\
2010-01-05	&   08:45:22.000014	&   M015	&   OFF	 &   \\
2010-01-05	&   08:45:24.000037	&   M012	&   ON	&   Kitchen Activity end \\
2010-01-05	&   08:45:26.000056	&   M023	&   OFF	 &   \\
\hline
\end{tabular}
\smallskip}
\caption{CASAS \cite{cook2012casas} Milan dataset anomaly}
\label{table:6}
\end{table}

\subsubsection{Evaluation Metrics}

\fix{Since HAR is a multiclass classification problem, researchers use metrics \cite{sokolova2009systematic} such as Accuracy, Precision, Recall, and F-Score to evaluate their algorithms \cite{park2018deep,singh2017convolutional, medina2018ensemble}. These metrics are defined by means of four features such as true Positive, true Negative , false Positive, and false Negative of class $C_i$. The F-score, also called the F1-score, is a measure of a model’s accuracy on a dataset. The F-score is a way of combining the Precision and Recall of the model, and it is defined as the harmonic mean of the model’s Precision and Recall. It should not be forgotten that real house datasets are mostly imbalanced in terms of class. In other words, some activities have more examples than others and are in minority. In an imbalanced dataset a minority class is harder to predict because there are few examples of this class, by definition. This means it is more challenging for a model to learn the characteristics of examples from this class, and to differentiate examples from this class from the majority class. Therefore it would be more appropriate to use metrics weighted by the class support of the dataset. Such as balanced Accuracy, weighted Precision, weighted Recall or weighted F-score \cite{fernandez2018learning, he2013imbalanced}.}

\subsection{Evaluation Process}

\subsubsection{Train / Test}

A first way to evaluate the algorithms is to divide the datasets into two distinct parts. One for training and the other for testing. It is generally chosen to use 70\% for training and 30\% for testing. Several researchers have chosen to adopt this method. Surong et al. \cite{yan2019using} have adopted this evaluation method in the application of real time activation recognition. In order to show the generalization of their approach, they chose to divide the datasets temporally into two equal parts. Then to re-divide each of these parts temporally into training and test datasets. They thus propose two sub-set of training and test. The advantage of this method is that it is usually preferable to the residual method and takes no longer to compute. Moreover, it does not allow to take into account the drift \cite{aminikhanghahi2019enhancing} of the activities. In addition it is always possible that the algorithm overfit on the test sets because the parameters have been adapted to optimal values. This approach does not guarantee a generalization of the algorithms.

\subsubsection{K-fold Cross Validation}

This is a wide approach used for model evaluation. It consists of dividing the dataset into K sub dataset, the value of K is often between 3 and 10. \hbox{K-1} dataset are selected for training and the remaining dataset for testing. The algorithm iterates until all the sub dataset is used for testing. The average of the training K scores is used to evaluate the generalization of the algorithm. It is usually customary that the data is mixed before being divided into K sub datasets in order to increase the generalization capability of the algorithms. However it is possible that some classes are not represented in the training or test sets. That's why some implementations propose that all classes are represented in tests as not training.

In the context of HAR in smart homes, this method is a good approach for classification of EW \cite{park2018deep,liciotti_lstm}. Indeed EW can be considered as independent and not temporally correlated. However it seems not relevant for sliding windows, especially if they have a strong overlap and the windows are distributed equally according to their class between the test and training set. The training and test sets would look too similar, which would increase the performance of the algorithms and would not allow it to generalize enough.

\subsubsection{Leave-One-Out Cross-Validation}
This is a special case of cross-validation where the number of folds equals the number of instances in the data set. Thus, the learning algorithm is applied once for each instance, using all other instances as a training set and using the selected instance as a single-item test set. 

Singh et al.\cite{singh2017recurent} and Medina-Quero et al. \cite{medina2018ensemble} used this validation method in a context of real-time HAR. In their experiments the dataset is divided into days. One day is used for testing while the other days are used for training. Each day becomes a test day in turn. This approach allows a large part of the dataset to be used for training. Allowing the algorithms to train on a wide variety of data. However the size of the test is not very significant and does not allow to demonstrate the generalization of the algorithm in the case of HAR in smart homes.

\subsubsection{Multi-Day Segment}
Aminikhanghahi et al. \cite{aminikhanghahi2019enhancing}  propose a validation method called Multi-Day Segment. This approach proposes to take into account the sequential nature of segmentation in a context of real-time HAR. Indeed, in this real-time context, each segment or window is temporally correlated. According to Aminikhanghahi et al, and as expressed above, cross validation would bias the results in this context. A possible solution would be to use the 2/3 training and 1/3 test partitioning as described above. However, this introduces the concept of drift into the data. Drift in terms of change in resident behavior would induce a big difference between the training and test set. 

To overcome these problems, the proposed method consists of dividing the dataset into 6 consecutive days. The first 4 days are used for training and the last 2 days are used for testing. This division into 6 day segments creates a rotation that allows to represent every day of the week in the training and test set. In order to make several folds, the beginning of the 6 day sequence is shifted 1 day forward at each fold. This approach allows to maintain the order of the data while avoiding the drift of the dataset.

\subsection{Outlines} 

Different validation methods for HAR in smart homes was reviewed in this section, Table \ref{tab:methods}. Depending on the problem being addressed, not all methods can be used to evaluate an algorithm. 

\begin{table}[ht!]
\centering
\resizebox{0.7\textwidth}{!}{%
\begin{tabularx}{\textwidth}{|Y|Y|Y|Y|Y|Y|Y|Y|Y|Y|}
\hline
\textbf{Ref}& \textbf{Train/Test spilt} & \textbf{K fold Cross validation} & \textbf{Leave One Out Cross Validation} & \textbf{Multi day segment} & \textbf{Respect Time order of activities} & \textbf{Sensitif to Data Drift problem} & \textbf{Real Time regognition} & \textbf{Ofiline recognition} & \textbf{Usable on small datasets} \\ \hline
\cite{yan2019using}& \Checkmark                       &                                  &                                         &                            & Yes                                       & Yes                                     & Yes                            & Yes                          & No                                \\ \hline
\cite{park2018deep,liciotti_lstm,gochoo2018unobtrusive}                                              &                           & \Checkmark                               &                                         &                            & No                                        & No                                      & No                             & Yes                          & No                                \\ \hline
\cite{singh2017recurent,medina2018ensemble,hamad2020joint,singh2017convolutional,wang2020activities} &                           &                                  & \Checkmark                                      &                            & Not necessarily                           & No                                      & Yes                            & Yes                          & Yes                               \\ \hline
\cite{aminikhanghahi2019enhancing}                                                                   &                           &                                  &                                         & \Checkmark                         & Yes                                       & No                                      & Yes                            & Yes                          & No                                \\ \hline
\end{tabularx}%
}
\caption{Summary of methods for evaluating activity recognition algorithms.}
\label{tab:methods}
\end{table}

In the case of offline HAR i.e. with EW or pre-segmented activity sequences, the K-fold cross-validation seems to be the most suitable. Provided that the time dependency between segments is not taken into account. Otherwise, it is preferable to use another method. 
\fix{The Leave-One-Out Cross-Validation approach is an alternative. It allows to process datasets containing few data. But the days are considered as independent. It is not possible to make a link between two different days e.g. a weekday or a weekend day.}
Aminikhanghahi et al. \cite{aminikhanghahi2019enhancing} have proposed a method to preserve the temporal dependence of the segments and avoid the problem of data drift induced by changes in the habits of the resident(s) over time.

In addition, the preprocessing of dataset data, the rebalancing, the removal of the ``Other'' class and the annotation of events affect the algorithms' performance. It is therefore important to take into account the evaluation method and the preprocessing performed, in order to judge the performance of the algorithm.  Moreover, classic metrics such as accuracy or F score may not be sufficient. It may be more judicious to use, metrics weighted by the number of representations of dataset classes such as dataset are unbalanced. Balanced accuracy, or F1 weighted score should be a better metric in this case \cite{fernandez2018learning, he2013imbalanced}.

A major problem in the area of HAR in smart homes is the lack of evaluation protocols. Establishing a uniform protocol according to the type of problem to be solved (real-time, offline) would speed up research in this field and allow a fairer comparison between the proposed algorithms and approaches.

\section{General Conclusion and Discussions}
In this article, we have highlighted the challenges of Human Activity Recognition in smart homes, some of which have particularities compared to other fields of HAR. We have proposed a taxonomy of the main components of a human activity recognition algorithm and reviewed the most promising solutions. To overcome the current issues, we point out the opportunities provided by new advances from other fields. 

\subsection{Comparison with Other HAR}
While human activity recognition algorithms have seen tremendous improvements for vision-based data owing to the rapid development of deep learning for image processing, human activity recognition using wearables and sensors on objects are also seeing significant improvements.  However, vision-based systems are seen by users as too intrusive as these systems could unveil too much private information, whereas wearables and sensors on objects require the daily instrumentation of the sensors on the body of subjects or their personal objects, ambient sensors could provide a solution to tackle this issue.

HAR in smart homes have seen recent advances owing to the development of recent deep learning algorithms for end-to-end classification such as convolutional neural networks. It also benefits from recent algorithms for sequence learning such as long-short term memory, but as with video processing, sequence learning still needs to be improved to both be able to deal with the vanishing gradient problem and to take into account the context of the sensor readings. The temporal dimension is incidentally a particularity of ambient sensor systems, as the data for a sparse and irregular time series. The irregular sampling in time has also been tackled with adapted windowing methods for data segmentation. In addition to the time windows used in other HAR fields, sensor event windows are also commonly used. The sparsity of the data of ambient sensors do not allow machine learning algorithms to take advantage of the redundancy of data over time, as in the case of videos where successive video frames are mostly similar. Moreover, whereas HAR in videos, the context of the human action can be seen in the images by the detection of his environment or objects of attention, the sparsity of the HAR in ambient sensors result in a high reliance in the past information to infer the context information. 

While HAR in ambient sensors have to face the problems of complex activities such as sequences of activities, concurrent activities or multi-occupant activities, or data drift, it also has to tackle specific unsolved problems such as the variability of data. Indeed, the data collected by sensors are even more sensitive to the house configuration, the choice of sensors and their localisation.  

\subsection{Taxonomy and Challenges}
To face its specific challenges and the challenges common to other systems, in our review, we introduced a taxonomy of the main components of a human activity recognition algorithm for real-use. \textbf{The three components we have pointed out are: classification, automatic feature extraction and time series analysis}. It needs to carry out a pattern recognition from raw data, thus requiring feature extraction. Moreover, the algorithm must integrate a time series analysis. 

While pattern recognition analysis and the feature extraction challenges seem to be well tackled by deep learning algorithms such as CNN, the sequence analysis parts have improved recently with the application of LSTM. Both approaches based on CNN and LSTM are reported to give equivalent performance levels and state-of-the-art developments are mostly based on either LSTM or convolutional deep learning. However the sequence analysis challenges still remain largely unsolved because of the impact of  the sparsity and irregularity of the data on context understanding and long-term reasoning. In particular, it makes the challenges of composite activities (sequences of activities), concurrent activities, multi-user activities recognition and data drift more difficult. The sparsity of the data also makes it more difficult to cope with the variability of the smart home data in its various settings. 

According to our analysis, the state of the art in HAR for ambient sensors are still far from ready to be deployed in real-use cases. To achieve this, the field must address the shortcomings of datasets, but needs also to standardise the evaluation metrics so as to reflect the requirements for a real-use deployment and to enable fair comparison between algorithms.

\subsection{Opportunities}
Moreover, we believe that recent advances in machine learning from other fields also offer opportunities for significant advances in HAR in smart homes. 

We advocate that the application of recent NLP techniques can bring advances in solving some of these challenges. Indeed, NLP also deploys methods of sequence analysis, and has also seen tremendous advances in the recent years. For instance, sparsity of the data can be alleviated by a better domain knowledge in the form of an emerging semantic. Thus taking inspiration from word encoding and language models, we can automatically introduce semantic knowledge between activities, as shown in the preliminary study \cite{Bouchabou2021I2WDLHAR}. Furthermore, a semantic encoding of the data will also help the system be more robust to unknown data as in the challenges of data drift or adaptation to changes, as it could be able to relate new data semantically to known data. 
Besides, the recent techniques for analysing long texts by inferring long-term context but also analysing the sequences of words and sentences, can serve as an inspiration to analyse sequences of activities or composite activities.

Lastly, we think that the unsolved problem of adaptation to changes of habits, users or sensor sets could soon find its solution in the current research on meta learning and interactive learning.

\subsection{Discussion}
In this review, we have pointed out the key elements for an efficient algorithm of human activity recognition in smart homes. We have also pointed out the most efficient methods, but also the remaining challenges and present opportunities. 
However the full deployment of smart home services, beyond the HAR algorithms, depend also on the development of the hardware systems and the acceptability and usability of these systems by final users.

For the hardware systems, the development of IoT devices with the improvement in the accuracy and autonomy along with the decrease in their cost will make them accessible to normal households. Despite cheaper sensors and actuators, it will not be realistic to provide all homes with a large set of sensors as in the current datasets, but real homes are not equipped as lavishly. smart home system thus need to optimise their hardware under constraints of budget, house configuration, number of inhabitants.... Smart home builder companies need to provide an adequate HAR hardware kit. To determine the minimal set of sensors, recently Bolleddula et al. \cite{bolleddula2020sensor}  used PCA to determine the most important sensors in a lavishly equipped smart home. This study is a first work to imagine a minimal sensors setup.

Finally, while IoT devices seem to be better accepted by users than cameras, there are still social barriers to the adoption of smart homes that need to be overcome \cite{balta2013social}. These require a trustworthy privacy-preserving data management but also reliable cyber-secure systems.

\textbf{Acknowledgement}: This work is partially supported by project VITAAL and is financed by Brest Metropole, the region of Brittany and the European Regional Development Fund (ERDF). This work was carried out within the context of a CIFRE agreement with the company Delta Dore in Bonemain 35270 France, managed by the National Association of Technical Research (ANRT) in France.

\vspace{6pt} 

\end{paracol}
\reftitle{References}


\externalbibliography{yes}
\bibliography{ref.bib}

%


\end{document}